\documentclass[11pt,a4paper]{article}
\pdfoutput=1
\usepackage{amsfonts,amssymb,epsfig,amsmath,enumitem,mathtools}
\usepackage{graphicx}
\usepackage[utf8]{inputenc}
\usepackage{cite}
\usepackage{hyperref,caption,subcaption}
\usepackage{xcolor,tikz}

\renewcommand{\thefootnote}{\fnsymbol{footnote}}

\allowdisplaybreaks

\newcommand{\nn}[0]{\nonumber}

\begin{document}

\makeatletter \@addtoreset{equation}{section} \makeatother
\renewcommand{\theequation}{\thesection.\arabic{equation}}
\renewcommand{\thefootnote}{\fnsymbol{footnote}}	

\begin{titlepage}
\begin{center}

\hfill {\tt QMUL-PH-20-11}\\
\hfill {\tt SNUTP20-001}\\
\hfill {\tt KIAS-P20020}

\vspace{1.5cm}

{\LARGE\bf AdS black holes and finite $N$ indices}

\vspace{1cm}

{\large Prarit Agarwal$^1$, Sunjin Choi$^2$, Joonho Kim$^3$, Seok Kim$^2$ 
and June Nahmgoong$^4$}

\vspace{1cm}
\textit{$^1$CRST and School of Physics and Astronomy, Queen Mary University of London, \\London E1 4NS, United Kingdom}\\
\vspace{0.2cm}
\textit{$^2$Department of Physics and Astronomy \& Center for
Theoretical Physics,\\Seoul National University, Seoul 08826, Korea.}\\
\vspace{0.2cm}
\textit{$^3$Institute for Advanced Study, Princeton, NJ 08540, USA.}\\
\vspace{0.2cm}
\textit{$^4$School of Physics, Korea Institute for Advanced Study, Seoul 02455, Korea.}\\

\vspace{1cm}

E-mails: {\tt  agarwalprarit@gmail.com, csj37100@snu.ac.kr, joonhokim@ias.edu, \\ skim@phya.snu.ac.kr, junenahmgoong@gmail.com}

\end{center}

\vspace{2.5cm}

\begin{abstract}
\normalsize

We study the index of 4d $\mathcal{N}=4$ Yang-Mills theory with
$U(N)$ gauge group, focussing on the physics of the dual BPS black holes in
$AdS_5\times S^5$. Certain aspects of these black holes can
be studied from finite $N$ indices with reasonably large $N^2$.
We make numerical studies of the index for $N\leq 6$, by expanding it up to
reasonably high orders in the fugacity. 
The entropy of the index agrees
very well with the Bekenstein-Hawking entropy of the dual black holes,
say at $N^2=25$ or $36$. Our data clarifies and supports the
recent ideas which allowed analytic studies of these black holes
from the index, such as the complex saddle points of the Legendre
transformation and the oscillating signs in the index. In particular,
the complex saddle points naturally explain the $\frac{1}{N}$-subleading
oscillating patterns of the index. We also illustrate the universality of
our ideas by studying a model given by the inverse of the 
MacMahon function.

\end{abstract}

\end{titlepage}

\renewcommand{\thefootnote}{\arabic{footnote}}

\setcounter{footnote}{0}

\setcounter{tocdepth}{2}
\tableofcontents

\section{Introduction and summary}

The superconformal index of large $N$ field theories 
\cite{Romelsberger:2005eg,Kinney:2005ej, Bhattacharya:2008zy} has recently 
received some attention \cite{Cabo-Bizet:2018ehj,Choi:2018hmj,Choi:2018vbz,Benini:2018ywd,Honda:2019cio,ArabiArdehali:2019tdm,Choi:2019miv,Kim:2019yrz,Cabo-Bizet:2019osg,Amariti:2019mgp,Kantor:2019lfo,Nahmgoong:2019hko,Lezcano:2019pae,Choi:2019zpz,Choi:2019dfu,ArabiArdehali:2019orz,Crichigno:2020ouj} as it successfully accounts for the thermodynamic entropy of AdS black holes.
In this paper, we want to present our numerical
study of the 4d $\mathcal{N}=4$ superconformal index, showing that some aspects 
of the BPS black holes in AdS$_5$ \cite{Gutowski:2004ez, Gutowski:2004yv, Kunduri:2006ek,Chong:2005da} 
can be investigated by numerically studying the index at finite $N$.
Our numerical data will also nontrivially support certain recent
ideas which enabled the analytic studies of these AdS black holes.

We define the Witten index of 4d $\mathcal{N}=4$  superconformal field theory on $S^3 \times S^1$  as \cite{Kinney:2005ej}
\begin{align}
    Z(\Delta_I, \omega_i) = \text{Tr}\left[(-1)^F e^{-\sum_{I=1}^3 \Delta_I Q_I - \sum_{i=1}^2 \omega_i J_i}\right]
\end{align}
with the constraint $\Delta_1 + \Delta_2 + \Delta_3 - \omega_1 - \omega_2 = 0$ on the chemical potentials. $Q_I$ with $I=1,2,3$ denote the $U(1)^3 \subset SO(6)$ R-charges of $\mathcal{N}=4$ superalgebra, and $J_i$ with $i=1,2$ denote the $U(1)^2 \subset SO(4)$ angular momenta on $S^3$. Only the BPS states with the energy $E = \sum_{I=1}^3 Q_I + \sum_{i=1}^2 J_i$ can contribute to the index.
Since the supersymmetric index is invariant under the continuous deformation of the gauge coupling, one can evaluate the index from the weakly interacting QFT. 
It can be done in a few steps. First, we obtain the following single-letter index \cite{Kinney:2005ej}
\begin{align}\label{letter-index}
    I_\text{single}(\Delta_I,\omega_i)
    = 1 - \frac{(1-e^{-\Delta_1})(1-e^{-\Delta_2})
    (1-e^{-\Delta_3})}{(1-e^{-\omega_1})(1-e^{-\omega_2})}
\end{align}
by counting all single-letter operators in the $\mathcal{N}=4$ vector multiplet that satisfy the above mentioned BPS energy condition. Next, we apply the Plethystic exponential to this index $I_\text{single}$ multiplied by the adjoint character $\chi_{\mathfrak{g}}(\mathbf{z})$ of the gauge algebra $\mathfrak{g}$, \cite{Kinney:2005ej}
\begin{align}
    \text{PE}[I_\text{single}(\Delta_I,\omega_i) \, \chi_{\mathfrak{g}}(z_a)] \equiv \exp\left[\sum_{n=1}^\infty \frac{I_\text{single}(n\Delta_I,n\omega_i) \, \chi_{\mathfrak{g}}(z_a^n)}{n}   \right] .
\end{align}
Finally, we project to the set of gauge invariant states by integrating over 
$z_a$ with the Haar measure of the gauge group.
The index of the 4d $\mathcal{N}=4$ theory with a gauge group corresponding to the Lie algebra $\mathfrak{g}$ reduces to a matrix model calculation giving the following integral \cite{Kinney:2005ej}:
\begin{align}
    Z_\mathfrak{g} = \oint d\mu_{\mathfrak{g}}(\mathbf{z}) \ {\rm PE}
    \Big[I_{\text{single}}(\Delta_I,\omega_i)\chi_{\mathfrak{g}}(\mathbf{z})\Big] \ .
    \label{eq:SuperconformalIndexFull}
\end{align}
Here
$d\mu_{\mathfrak{g}}(\mathbf{z})$ is the Haar measure of $\mathfrak{g}$. Explicitly, it can be written as
\begin{align}
   \oint d\mu_{\mathfrak{g}}(\mathbf{z}) = \frac{1}{(2 \pi i)^r} \frac{1}{|W|} \oint_{|z_1|=1} \ldots \oint_{|z_r|=1} 
   \frac{dz_1 \ldots dz_r}{z_1\ldots z_r} 
   \prod_{\alpha \in \Delta} (1-{\mathbf{z}}^{\alpha}) \ ,
    \label{eq:HaarMeasureFull}
\end{align}
where $W$ is the Weyl group of $\mathfrak{g}$, $r$ is the rank, $z_a$ is the fugacity corresponding to its $a$-th Cartan generator and $\Delta$ is the set of its roots.
It turns out that for numerical purposes it is more efficient to use a slightly modified definition of the Haar measure given by restricting the product in \eqref{eq:HaarMeasureFull} to only the positive roots of $\mathfrak{g}$ \cite{Hanany:2008sb}:
\begin{align}
    \oint d\mu_{\mathfrak{g}}(\mathbf{z}) = \frac{1}{(2 \pi i)^r}  \oint_{|z_1|=1} \ldots \oint_{|z_r|=1} \frac{dz_1 \ldots dz_r}{z_1\ldots z_r} 
    \prod_{\alpha \in \Delta^+} (1-{\mathbf{z}}^{\alpha}) \ .
\end{align}
This helps by removing the need to normalize the integral by the order of the Weyl group. From (\ref{eq:SuperconformalIndexFull}) and
(\ref{letter-index}), $Z(\Delta_I,\omega_i)$ is invariant under
$2\pi i$ shift of each of $\Delta_I,\omega_i$. So one can equivalently study
the index at the surface $\sum_I\Delta_I-\sum_i\omega_i=2\pi i \mathbb{Z}$.
Below, we shall often choose the right hand side to be $2\pi i$.

For our purposes it suffices to consider a special unrefined 
case of the above integral by setting 
$e^{-\Delta_1}=e^{-\Delta_2}=e^{-\Delta_3}\equiv e^{-\Delta}$,
$e^{-\omega_1}=e^{-\omega_2}\equiv e^{-\omega}$.
If one Legendre transforms to the microcanonical ensemble at macroscopic charges,
this amounts to taking equal charges and equal angular momenta, $Q_1 = Q_2 = Q_3 = Q$ and $J_1 = J_2 = J$. From $3\Delta-2\omega=2\pi i\mathbb{Z}$,
one can set $x^2=e^{-\Delta}$, $x^3=e^{-\omega}$ for certain $x$.
The fugacity $x$ is now conjugate to the charge $j \equiv 6(Q+J)$.
The expression in \eqref{eq:SuperconformalIndexFull} then becomes
\begin{align}
   Z_\mathfrak{g} =  \oint d\mu_{\mathfrak{g}}(\mathbf{z}) \ {\rm PE}
   \Big[\left(1-\frac{(1-x^2)^3}{(1-x^3)^2}\right)
   \chi_{\mathfrak{g}}(\mathbf{z})\Big] \ .
    \label{eq:SuperconformalIndexReduced}
\end{align}

The resulting index can be expanded as
\begin{equation}
  Z=\sum_{j=0}^\infty \Omega_j x^j
\end{equation}
where $j\equiv 6(Q+J)$ and $\Omega_j$ are integers which count the
number of BPS states (with $-1$ factor for fermions). For $U(N)$ gauge group,
we shall study this index at $N=2,3,4,5,6$, by computing the coefficients of
the fugacity expansion in $x$ up to fairly high orders, till $\mathcal{O}(x^{100})$
for $N\leq 5$, and till $\mathcal{O}(x^{70})$ for $N=6$. Naively, finite $N$
indices will be irrelevant for studying emergent gravitational phenomena
expected in the large $N$ limit. In particular, one would like to study the
large $N$ limit of $\Omega_j$ when $j$ is of order $N^2\gg 1$.
In this limit, black hole like degeneracy will grow like
$\log|\Omega_i|\sim N^2$ when $j\sim N^2$. Our starting point is that,
in practice, taking $N=5$ or $6$ has already large enough $N^2$, so that
we can hope to see the black hole like exponential growth of $\Omega_j$
quite convincingly. In fact, plugging in $N^2=25$ or $36$ to
the geometric Bekenstein-Hawking entropy formula for the known AdS$_5$
black holes, we shall find very good agreements with the field theory
calculus of $\log|\Omega_j|$. In non-Abelian gauge theories, how small 
$\frac{1}{N}$ should be at finite $N$ to exhibit large $N$ behaviors 
depends on the type of physics one is
interested in. So not too surprisingly, our finite $N$ approach does not
clearly see certain types of black holes. For instance,
we empirically find that the charge range for the so-called
`small black holes' is not clearly resolved in our finite $N$ discretized
analysis. (See section 2 for more explanations.) The detailed physics
that can be learned is outlined below, and will be elaborated more in section 3.

Our finite (but reasonably large) $N$ calculus
reveals various interesting structures which shed more concrete
lights on the recent analytic studies of these black holes. 
After computing the large $N$ free energy
$\log Z$ as a function of chemical potentials $\Delta_I,\omega_i$,
one makes a Legendre transformation to the microcanonical ensemble
to compute the entropy. Legendre transformation is a saddle point approximation
of the inverse Laplace transformation
\begin{equation}\label{inverse-Laplace}
  \Omega_j=\oint\frac{dx}{x} x^{-j}Z(x)
\end{equation}
at macroscopic charge $j$. (The formula can be generalized to refined
$\Delta_I,\omega_i$, but we present the above unrefined formula for simplicity.)
The fact is that the dominant saddle point values $x_\ast$ of
$x$ (or $\Delta_I,\omega_i$) are
complex, at real $j$ (or $Q_I,J_i$). The naively computed saddle point
value of the integral,
$\Omega_j(x_\ast)\equiv e^{S(j)}$, at real positive $j$ is therefore complex.
Somewhat surprisingly, this simple fact apparently seems to have confused 
many people,
leading to a number of ad hoc prescriptions and interpretations on
how to extract the correct physics out of this result. We stick to the
natural interpretation of \cite{Choi:2018hmj,Choi:2018vbz,Choi:2019miv} and find
extremely nontrivial evidences supporting it from our numerical studies.
We think this will confirm our interpretation to be the canonical 
picture, which goes as follows.
From the unitarity of the underlying QFT, it is always guaranteed that
one can find the complex conjugate saddle point $\bar{x}_\ast$
for any complex $x_\ast$. The conjugate saddle point value is given by
$\Omega_j(\bar{x}_\ast)=
e^{\overline{S(j)}}$. Adding the two equally dominant contributions,
one obtains
\begin{equation}
  \Omega_j\sim\Omega_j(x_\ast)+\Omega_j(\bar{x}_\ast)
  \sim \exp\left[{\rm Re}(S(j))+\cdots\right]
  \cos\left[{\rm Im}(S(j))+\cdots\right]\ ,
\end{equation}
where $\cdots$ denote possible subleading corrections
at large $N^2$ and large $j$. (Note that ${\rm Re}(S(j))$ and
${\rm Im}(S(j))$ scale like $N^2$.)
As will be manifest from our data in the next
section, the integers $\Omega_j$ at macroscopic $j$ grow exponentially
fast to account for the dual black holes, but come with possible minus signs
at certain $j$'s. Namely, $\Omega_j$ as a function of (quantized) $j$
oscillates between positive and negative integers as $j$ changes. However,
the macroscopic Legendre transformation calculus is not sensitive to the precise
quantized nature of $j$ and $\Omega_j$. Therefore, the best one can expect
to see from this calculus is an exponentially growing envelope function,
which is provided by $e^{{\rm Re}(S(j))}$, multiplied by a factor
which oscillates between $+1$ and $-1$, which is provided by
$\cos\left[{\rm Im}(S(j))+\cdots\right]$ in the above expression.

Our numerical calculus will justify this interpretation. Firstly,
the computed entropy $\log |\Omega_j|$ from the integers $\Omega_j$
indeed takes the form of
\begin{equation}\label{entropy-subleading}
  {\rm Re}(S(j))+\log\left[\cos({\rm Im}(S(j)+\cdots)\right]\ ,
\end{equation}
where ${\rm Re}[S(j)]$ and ${\rm Im}[S(j)]$ are those computed recently
from the index using various analytic methods (in the large $N$ and/or
large charge limit). Furthermore, more importantly, investigating the overall
signs in $\Omega_j$
from our numerical calculus, the sign oscillating pattern is also determined
by the sign oscillation of $\cos\left({\rm Im}(S(j))+\cdots\right)$,
upon fitting a constant $\mathcal{O}(1)$ phase shift in `$\cdots$' that
has not yet been computed by any analytic methods. Therefore, a precise
interpretation is given to ${\rm Im}(S(j))$,
as containing the overall sign information of $\Omega_j$.

While comparing our numerically computed $\log |\Omega_j|$ with
(\ref{entropy-subleading}), confirming the appearance of the second term
is nontrivial. This is because,
while the first term is proportional to $N^2$, the second term is typically
subleading because the macroscopic quantity ${\rm Im}(S(j))\sim N^2$
is inside the cosine function. To detect the second term, it is 
crucial to make a precision computation of the index which sees this
`$\frac{1}{N}$ corrections.' Our finite $N$ indices (say at $N=5,6$) provide
a perfect setup to confirm such structures, as these values of $N^2$ are
large enough to provide a large $N$ hierarchy to various contributions to
the entropy, while not being too large so that the subleading corrections
are visible. We think our numerical support to the formula (\ref{entropy-subleading})
is compelling. See section 3 for the details.

The interpretations outlined above appear to be universal,
which may appear in any index-like generating functions that have negative
integer coefficients at various orders. We illustrate that this is actually
the case, by studying in detail the inverse of the MacMahon function
\begin{equation}
  f(x)\equiv\prod_{n=1}^\infty (1-x^n)^n=
  \sum_{j=0}^\infty \Omega_j x^j
  =1-x-2x^2-x^3+0x^4+4x^5+4x^6+7x^7+3x^8-2x^{9}-9x^{10}-17x^{11}-\cdots\ .
\end{equation}
At large $j$, one can analytically compute the macroscopic entropy
given by $\log |\Omega_j|\sim \frac{3}{4}\left[2\zeta(3)j^2
\right]^{\frac{1}{3}}+\cdots$, where
`$\cdots$' denotes small $\frac{1}{j}$ corrections which can be
concretely computed to any desired accuracy. On the other hand,
$\Omega_j$ exhibits a characteristic oscillation between positive and
negative integers. We shall illustrate that this is precisely realized
in the Legendre transformation as the complex saddle points, where a
formula like (\ref{entropy-subleading}) will provide a perfect match.
As we can explicitly compute the $\frac{1}{j}$ corrections to high orders,
including the finite phase shifts in the second term of (\ref{entropy-subleading}),
our interpretation can be tested to very high accuracy in this model.

The remaining part of this paper is organized as follows.
Section 2 summarizes our numerical results for the integers $\Omega_j$.
We also explain some salient structures of the series $\Omega_j$,
and also provide a comparison with the Bekenstein-Hawking entropy of
black holes. In section 3, we take a closer look at the structures of
$\Omega_j$ and the $\frac{1}{N}$ correction, and provide various
interpretations and discussions.

\section{Numerical study of the $\mathcal{N}=4$ index}
\label{sec:index}

\begin{figure}
  \begin{subfigure}{\textwidth}
    \centering
    \includegraphics[height=8cm]{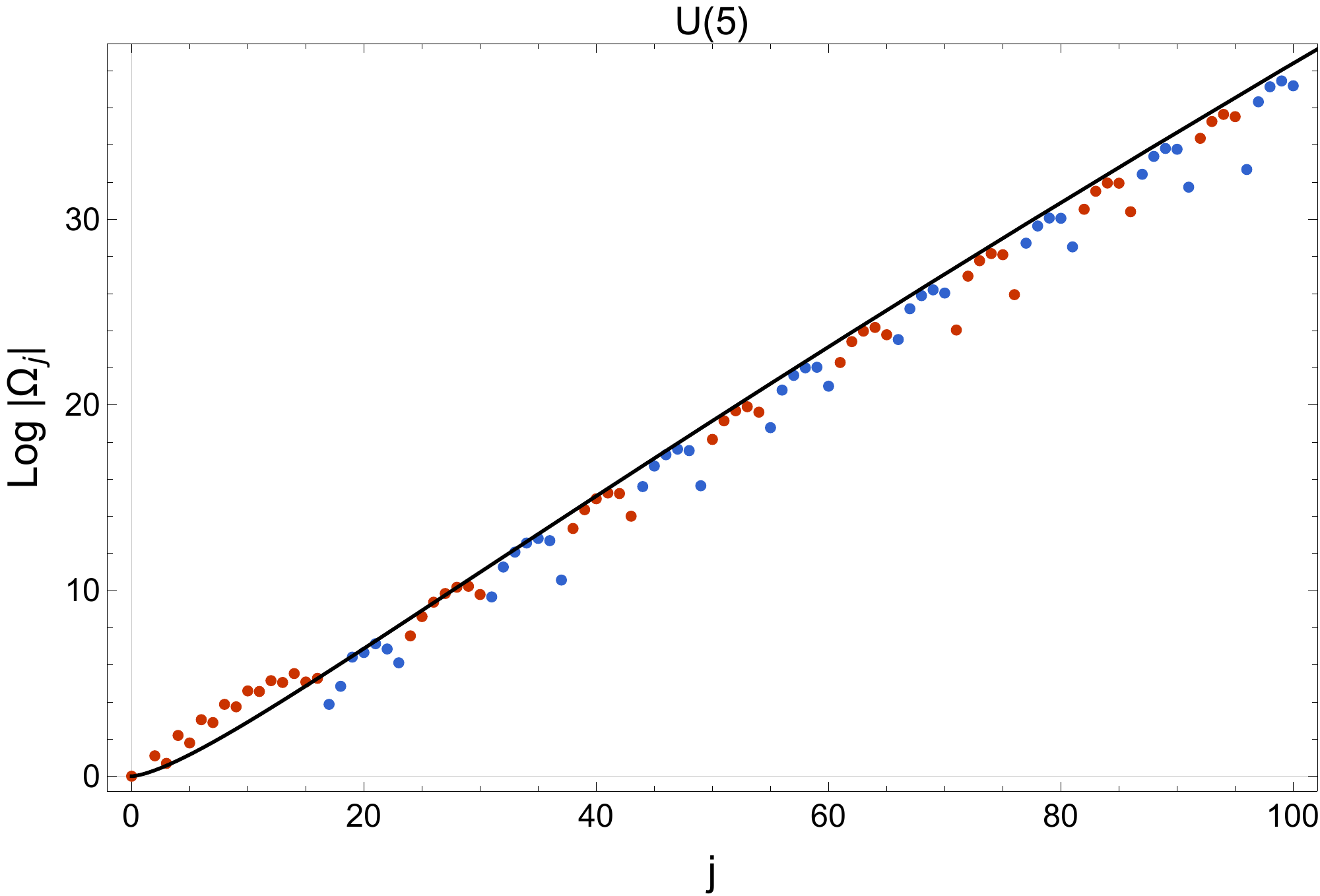}
    \caption{$N=5$}
    \label{fig:U5}
  \end{subfigure}
    \begin{subfigure}{\textwidth}
    \centering
    \includegraphics[height=8cm]{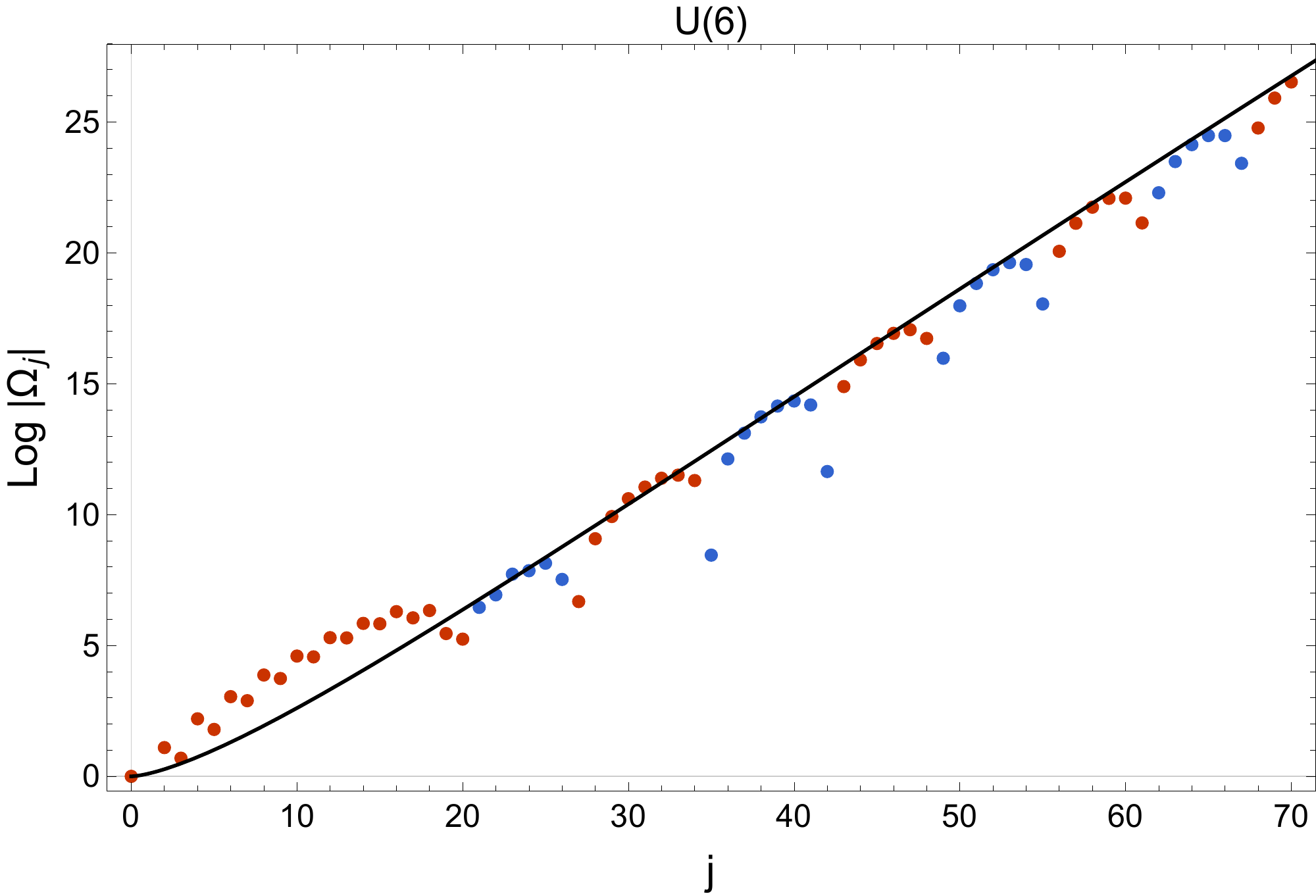}
    \caption{$N=6$}
    \label{fig:U6}
  \end{subfigure}
    \caption{Plots of $\log |\Omega_j|$ for $U(5)$ and $U(6)$ indices.
    The colors of the points encode the sign of $(-1)^j\Omega_j$: red being
    positive and blue being negative. ${\rm Re}(S(j))$ computed from the black hole
    entropy function is the Bekenstein-Hawking entropy, given by the curve drawn
    with a solid black line.}
    \label{fig:U5U6}
\end{figure}

We now specialize to the case of 4d $\mathcal{N}=4$ theories with a $U(N)$ gauge group. We would like to probe the regime
\begin{align}
    Q_I, J_i \sim N^2 \gg 1 \ .
\end{align}
However, the last inequality will be reasonably met by trying to take $N^2$
and charges to be as large as possible within our computational capability.
We expand the index in $x$ (as introduced in section 1), perform the integral over
$N$ variables on computer, to obtain various coefficients of
\begin{align}
    Z_{U(N)} = \sum_{j=0}^\infty \Omega_j x^{j} \qquad \text{with} \qquad j \equiv 6(Q+J) \ .
\end{align}
This is a straightforward exercise, with the main impediment coming from the availability of sufficient computing power.
The computational-complexity of the integral grows extremely quickly as the rank of the gauge group increases. We were able to explicitly evaluate the above integral for $2 \leq N\leq5$ up to $\mathcal{O}(x^{100})$, as given in \eqref{eq:u2index}--\eqref{eq:u5index}. For $U(6)$ we evaluated it up to $\mathcal{O}(x^{70})$. The explicit expression of the $U(6)$ index is given by:
\begin{align}
    \label{eq:u6index}
    Z_{U(6)} &=
    1 + 3 x^{2} - 2 x^{3} + 9 x^4 - 6 x^5 + 21 x^6 - 18 x^7 + 48 x^8 - 42 x^9 + 99 x^{10}
    - 96 x^{11} \\&
    + 200 x^{12} - 198 x^{13} + 345 x^{14} - 340 x^{15} + 540 x^{16} - 426 x^{17} + 564 x^{18} - 234 x^{19} \nonumber\\&
    + 189 x^{20} +
 636 x^{21} - 1026 x^{22} + 2262 x^{23} - 2583 x^{24} + 3438 x^{25} -
 1851 x^{26} \nonumber\\&
 - 794 x^{27} + 8757 x^{28} - 20460 x^{29} + 40398 x^{30} -
 63054 x^{31} + 88401 x^{32} \nonumber\\&
 - 99388 x^{33} + 80856 x^{34} + 4680 x^{35} -
 184576 x^{36} + 494910 x^{37} - 920943 x^{38} \nonumber\\&
 + 1392360 x^{39} -
 1690101 x^{40} + 1451568 x^{41} - 114147 x^{42} - 2931498 x^{43} +
 8129358 x^{44} \nonumber\\&
 - 15183836 x^{45} + 22398435 x^{46} - 25748382 x^{47} +
 18439724 x^{48} + 8645112 x^{49} \nonumber\\&
 - 64166661 x^{50} + 150570130 x^{51} -
 254339973 x^{52} + 334069536 x^{53} - 310532838 x^{54}\nonumber \\&
 + 68770386 x^{55} +
 514459605 x^{56} - 1501534768 x^{57} + 2775637323 x^{58} -
 3887229606 x^{59} \nonumber\\&
 + 3923925613 x^{60} - 1520426502 x^{61} -
 4814089191 x^{62} + 15863550944 x^{63} \nonumber \\&
 - 30282658596 x^{64}+ 42802285428 x^{65} - 42817602705 x^{66} + 14831924490 x^{67} \nonumber\\&
 + 57170104014 x^{68} - 179436305580 x^{69} + 331894244529 x^{70}
 +\mathcal{O}(x^{71})\ .\nonumber
\end{align}
It was pointed out in \cite{Choi:2018vbz} that the alternation of $\pm$ signs
of $\Omega_j$ demands special care when one attempts to extract it out at large $j$
using Legendre transformation. These sign alternations are generic: they also
happen at lower $N$'s. See the results in the appendix A.
We shall later observe more organized patterns of the sign alternations,
as will be explained in section 3.
Here, we simply note that the absolute degeneracy $|\Omega_j|$ indeed grows
very fast at large $j$. For instance, one finds $|\Omega_{70}|\sim 3.3\times 10^{11}$
at $N=6$, and $|\Omega_{100}|\sim 1.4\times 10^{16}$ at $N=5$. We will see shortly
that $\Omega_j$ grows quantitatively like the black hole entropy even at $N=5,6$.
See Fig.~\ref{fig:U5U6} for $\log |\Omega_j|$ and the signs of $\Omega_j$ at
$N=5,6$.

We want to compare our indices at reasonably large $N$
with the spectra in the gravitational dual. At low energies, the BPS
spectrum can be computed from the gas of gravitons \cite{Kinney:2005ej}.
A BPS graviton particle corresponds to a particular single trace BPS
operator in the QFT dual. It is a valid approach when the energy $E$ satisfies
$E\ll N$. In this limit, the BPS multi-graviton states correspond to
multi-trace operators obtained by multiplying the above mentioned single
trace operators, where one does not have to consider trace relations.
As the energy grows, the finite $N$ effects of these graviton states
have been studied in some detail in the BPS sector. The 
trace relations will start to enter from an energy of order $N$,
reducing the number of independent operators than the naive multi-particle
spectrum beyond this threshold. To see how this picture is reflected in
our $\Omega_j$'s, we first consider the index over BPS gravitons
given by \cite{Kinney:2005ej}
\begin{align}
    \label{eq:Graviton}
    \sum_{j=0}^\infty \Omega_j^g x^j \equiv \prod_{n=1}^\infty \frac{(1-x^{3n})^2}{(1-x^{2n})^3}\ .
\end{align}
Comparing our $\Omega_j$ and $\Omega_j^g$, one finds that
$\Omega_j=\Omega_j^g$ holds for $j\leq 2N+1$. This can be seen exactly
for all $N=2,\cdots,6$, and presumably holds exactly for other values of $N$.
Slightly beyond this point, $j\gtrsim 2N+1$,
$|\Omega_j|$ is smaller than $|\Omega_j^g|$ for a certain while.
So $j=2N+1$ is naturally interpreted as
the threshold where the trace relation starts to reduce the BPS states.

Now we consider the regime in which $j$ is substantially larger than this
threshold, so that the resulting $|\Omega_j|$ cannot be explained from
$|\Omega_j^g|$ with the trace relation reduction.
($|\Omega_j|$ becomes bigger than $|\Omega_j^g|$ for sufficiently large $j$.)
Eventually we enter a region with $j\sim N^2$,
whose gravitational dual description will be the BPS black holes in AdS.
To provide the comparison with the Bekenstein-Hawking entropy of these
black holes, let us first explain the entropy function approach
to understand its structures in a simple manner \cite{Hosseini:2017mds}.
We present the results in the version which only keeps one fugacity
$x$ \cite{Choi:2018hmj,Choi:2018vbz}.
The entropy function we shall discuss assumes the convention
$3\Delta-2\omega=2\pi i$. Then
$x^2=e^{-\Delta}$, $x^3=e^{-\omega}$ can be solved as
\begin{equation}
  x=e^{-\frac{\omega}{3}+\frac{2\pi i}{3}}=-e^{-\frac{\Delta}{2}}\ .
\end{equation}
In this setup, consider the following entropy function of $j$ and $\omega$:
\begin{align}
    \label{eq:EntropyFtn}
    S(\omega,j) = \frac{N^2\Delta^3}{2\omega^2}+\frac{\omega-2\pi i}{3}j
    =\frac{N^2}{2\omega^2} \left(\frac{2\pi i + 2\omega}{3}\right)^3
    +\frac{\omega-2\pi i}{3}j\ .
\end{align}
The first term on the right hand side 
originates from $\log Z$ in the grand canonical ensemble,
and the second term is the Legendre transformation factor, whose
exponential becomes $x^{-j}$ of (\ref{inverse-Laplace}).
At fixed charge $j>0$, one extremizes $S(\omega,j)$ with $\omega$.
This yields a cubic equation in $\omega$, which yields three different solutions
$\omega_\ast$. Among these three, we take the one which yields
maximal ${\rm Re}(S(j))>0$ where $S(j)\equiv S(\omega_\ast,j)$.
At this solution, one finds
\begin{align}
    \omega_\ast &= -\xi \sqrt{\frac{3\pi + 3\xi}{\pi-3\xi}} + i\xi \\
    j &= -\frac{N^2}{9}\frac{(\pi - 2\xi)^2(\pi+ \xi)}{\xi^3}\nn \\
    \log{Z} &= +\frac{N^2}{18 } \frac{\pi ^3-9 \pi  \xi^2-8 \xi ^3}{\xi^2} \sqrt{\frac{\pi +\xi}{3 \pi -9 \xi }} - i\frac{N^2}{54}\frac{(\pi-8\xi)(\pi+\xi)^2}{\xi^2}\ , \nn
\end{align}
where $\xi$ is a real number satisfying $-\pi<\xi<0$. It parametrizes
the imaginary part of $\omega$, and is a monotonically 
increasing function of $j$ implicitly given
by the second line. Inserting this value back to
$S(\omega,j)$, one obtains $S(j)$ given by
\begin{align}\label{BH-entropy}
    \text{Re}(S(j)) & = \frac{N^2}{6}\frac{\pi    (\pi ^2-2 \pi  \xi-3 \xi ^2 )}{ \xi ^2}\sqrt{\frac{\pi+\xi  }{3 \pi -9 \xi }}
    \\
    \text{Im}(S(j)) & =
    -\frac{N^2}{18}\frac{\pi(\pi -5 \xi ) (\xi +\pi )}{ \xi ^2}
     -\frac{2\pi}{3}j\nn
\end{align}
where the relation $j(\xi)$ is assumed. The fact is that
${\rm Re}(S(j))$ is precisely the Bekenstein-Hawking entropy of
the BPS AdS black holes of \cite{Gutowski:2004ez, Gutowski:2004yv, Kunduri:2006ek} at
$Q\equiv Q_1=Q_2=Q_3$ and $J\equiv J_1=J_2$. More precisely,
\cite{Gutowski:2004ez, Gutowski:2004yv, Kunduri:2006ek} found black hole solutions carrying two
charges $Q$, $J$, depending on only one independent parameter. The entropy
is a function of this parameter, which is in one to one correspondence with
$j\equiv 6(Q+J)$. Therefore, expressing the one-parameter
Bekenstein-Hawking entropy in terms of $j$, one obtains
the above ${\rm Re}(S(j))$.
Here, $N^2$ in the gravity side is related to the inverse Newton constant
$G_5^{-1}$ of the 5d gravity as $N^2=\frac{\pi\ell^3}{2G_5}$, where
$\ell$ is the radius of AdS$_5$.

The classical gravity description will be reliable at small
enough Newton constant, i.e. $N^2\gg 1$. To compare with our numerical
results at $N=5,6$, we plug in $N^2=25$ or $36$ to (\ref{BH-entropy})
expecting that $N^2$ is reasonably large. In Fig.~\ref{fig:U5U6}, we have drawn
these ${\rm Re}(S(j))$ by the black solid lines. At large enough
charge $j$ (especially for $U(5)$ where we could do numerical calculations
for larger charges), this agrees very well with the numerically computed
entropy $\log |\Omega_j|$ of the index. There appear intriguing
oscillations of our numerical $\log|\Omega_j|$,
which appear to be subleading in $\frac{1}{N}$ at large enough charges.
We shall comment on these subleading fluctuations in the next section.

Similar plots are shown for lower $N$ in the appendix. Of course,
inserting the finite values of $N^2$ to (\ref{BH-entropy}) becomes
less meaningful for those lower values. As one can see from these figures,
the numerical $\log|\Omega_j|$ and ${\rm Re}(S(j))$ do not agree that
well for $N=2$ or $N=3$. Here we note that, although $S(j)$ of (\ref{BH-entropy})
is introduced here as the entropy function for the black hole, valid at
$N^2\gg 1$, it has been shown \cite{Choi:2018hmj} that
(\ref{eq:EntropyFtn}) and (\ref{BH-entropy}) are true
at any finite $N^2$ when $\omega$ becomes small (or equivalently, when
$j\gg N^2$). This is called the `Cardy limit' of higher dimensional SCFTs
in the recent literature. In this case, (\ref{BH-entropy}) and
(\ref{eq:EntropyFtn}) have been derived from the field theory side for
any value of $N$. As one can see gaps between $\log |\Omega_j|$ and
${\rm Re}(S(j))$ for $N=2,3$ in Figs. \ref{fig:U2} and \ref{fig:U3} in
appendix A, it appears that the charge $j=100$ has not yet reached the Cardy regime.

We can also try to characterize which kinds of black holes are well
described by our numerical data, and which kinds are not well visible.
In AdS, one can classify black holes into `small black holes' and
`large black holes' depending on various (closely related) criteria.
The classification was originally made for
AdS Schwarzschild black holes. However, similar notion exists for our
BPS black holes by the charge playing the role of energy, and
the inverse chemical potential playing the role of temperature.
The most intuitive way to distinguish the AdS black holes is whether
the `size' of the black hole is smaller than the AdS radius $\ell$,
or larger than it. To make it more precise, consider the temperature $T$
of the black hole given by $\frac{1}{T}=\frac{dS(E)}{dE}$.
For our BPS black holes, ${\rm Re}(\omega)$, $j$, ${\rm Re}(S(j))$ play
the role of $T^{-1}$, $E$, $S(E)$ respectively. They satisfy
the analogous relation
\begin{equation}
  \frac{1}{3}{\rm Re}(\omega)=\frac{d[{\rm Re}S(j)]}{dj}\ .
\end{equation}
Now consider taking the second derivative with energy (or $j$),
\begin{equation}
  \frac{dT^{-1}(E)}{dE}=\frac{d^2S(E)}{dE^2}\ \ ,\ \
  \frac{1}{3}\frac{d{\rm Re}(\omega(j))}{dj}=\frac{d^2[{\rm Re}S(j)]}{dj^2}\ ,
\end{equation}
where the first and second expressions apply for Schwarzschild black holes
and our BPS black holes. The negativity of these expressions implies
that the black holes are stable in the canonical and grand canonical
ensemble, respectively, due to the heat capacity or susceptibility
being positive. We call these black holes `large black holes.'
They are characterized by the entropy being a convex function of
$E$ or $j$. Our BPS black holes are in the large black hole branch for
$j> j_0\equiv \frac{(5+3\sqrt{3})N^2}{9}$ (or $-\frac{\pi}{\sqrt{3}}<\xi<0$).
On the other hand,
for $j<j_0$ (or $-\pi<\xi<-\frac{\pi}{\sqrt{3}}$),
the curve $S(j)$ is concave and one is in the small black hole branch.
As one sees from the black curves in Fig.~\ref{fig:U5U6},
the visibly concave region is at so small charges,
that they are essentially overlapping with the region
$j\leq 2N+1$ in which the graviton description is good. Namely,
we find that the small black hole branch squeezed by
the graviton region from the left and $j_0$ from the right 
is not clearly visible from our finite $N$ indices.
At large enough $N$, the two charge scales
$j\sim 2N+1$ and $j\sim\ j_0$ will be given enough hierarchy to
allow a visible small black hole region. However, our finite $N$
index does not seem to have large enough $N$ to make this region
clearly visible. Indeed, this can be clearly seen from our numerical plots
in Fig.~\ref{fig:U5U6}. In the small black hole region,
$S(j)$ will increase very fast in $j$. However, our numerical
$\log |\Omega_j|$ does not manifestly exhibit such an inflating region.
It will be interesting to compute $\Omega_j$'s for larger $N$'s to
see this region.

So far, we explained how to compare our $\log |\Omega_j|$ with
${\rm Re}(S(j))$ of the dual black holes.
There is other interesting information that one can get from
our numerical data, concerning ${\rm Im}(S(j))$, the signs of
$\Omega_j$, and the subleading oscillations that one sees in the
figures. These will be discussed in more detail in the next section.

\section{Interpretations and discussions}

In this section, we discuss more detailed information encoded in our
numerical $\Omega_j$, and relate it to the interpretations made on
(\ref{eq:EntropyFtn}).

We first study the signs of $\Omega_j$. The pattern of the signs
visible in the series $Z(x)=\sum_j\Omega_j x^j$ apparently looks very
complicated. However, one observes simplifications upon inserting
$x\rightarrow -x$:
\begin{equation}
  Z(-x)=\sum_{j}(-1)^j\Omega_j x^j\ .
\end{equation}
The signs of $(-1)^j\Omega_j$ are shown in Fig.~\ref{fig:U5U6} and
also in the figures of appendix A by the colors of the dots.
After this substitution, one finds that the sign change pattern is
correlated to the subleading oscillation pattern of
$\log |\Omega_j|$. Namely, the sign changes only at the local minima
of the oscillation.

\begin{figure}
  \begin{subfigure}{\textwidth}
    \centering
    \includegraphics[height=8cm]{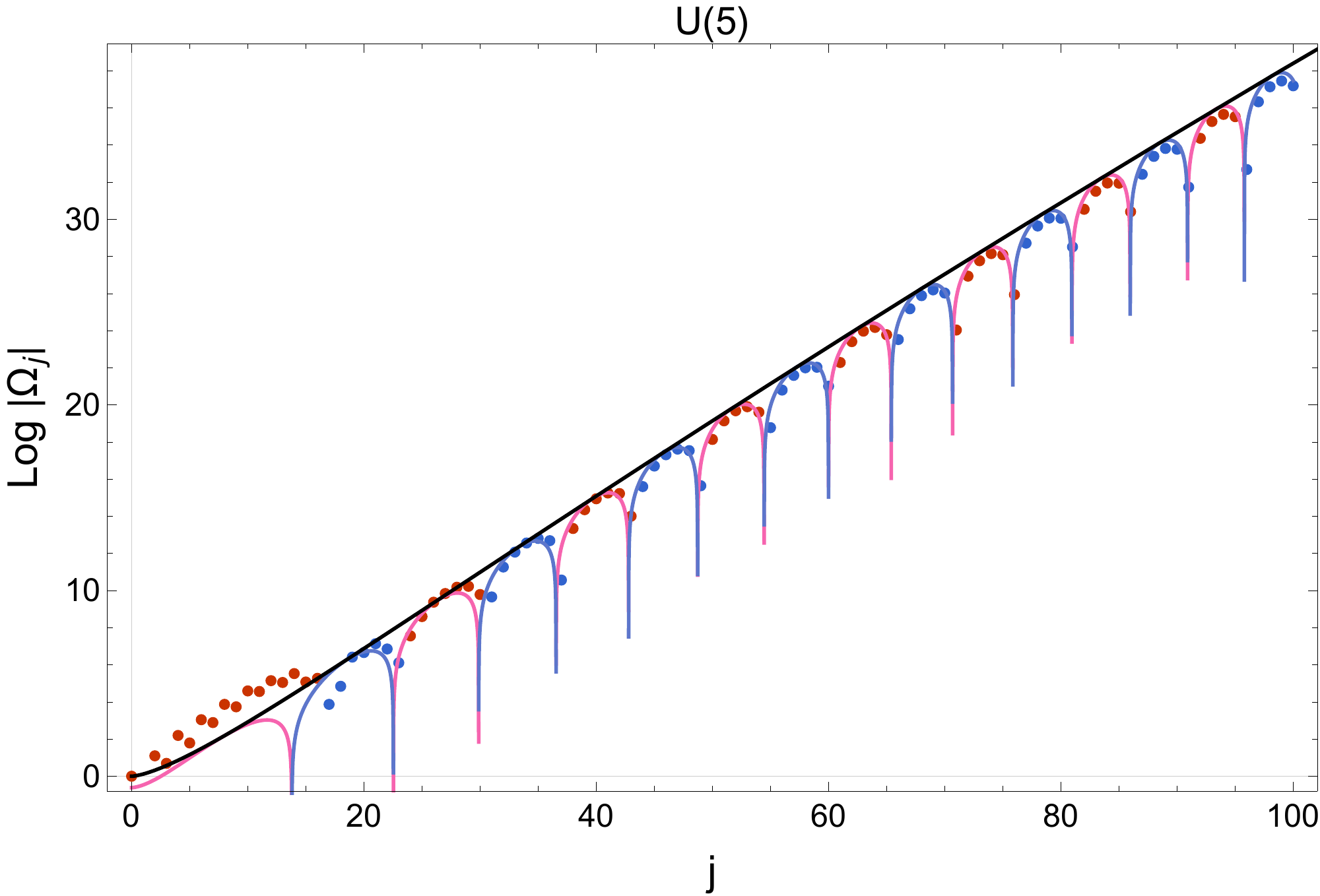}
    \caption{$N=5$}
    \label{fig:U5bump}
  \end{subfigure}
    \begin{subfigure}{\textwidth}
    \centering
    \includegraphics[height=8cm]{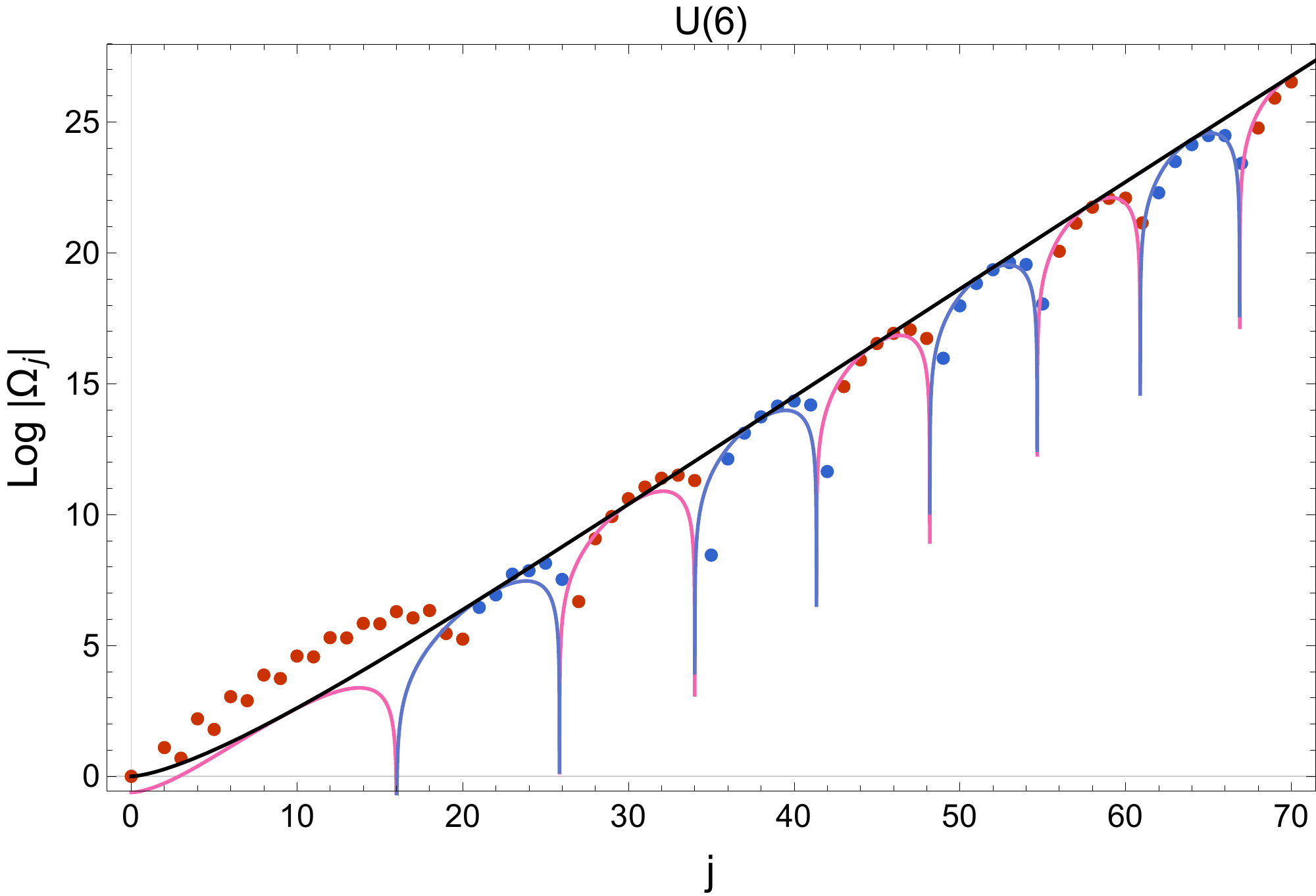}
    \caption{$N=6$}
    \label{fig:U6bump}
  \end{subfigure}
    \caption{Same plots as Fig.~\ref{fig:U5U6}, with the extra
red/blue curves for 
${\rm Re}(S(j))+\log\left|\cos\left[{\rm Im}(S(j))+\pi j+\eta
   \right]\right|$. 
A subleading constant $\eta$ is empirically tuned to
    $\eta\approx -1$ to minimize the overall off-phase behaviors. 
    The red and blue colors of the curves
    denote $\cos[{\rm Im}(S(j))+\pi j+\eta]\gtrless 0$, respectively.}
    \label{fig:U5U6bump}
\end{figure}
At this point, we revisit the interpretation of complex $S(j)$ at the
saddle point of the Legendre transformation at macroscopic charges,
that we outlined in section 1. The interpretation asserts that
the sign of $\cos[{\rm Im}(S(j))+\cdots]$ equals
the sign of the integers $\Omega_j$. Since we have observed very
simple sign oscillation patterns of our data $(-1)^j\Omega_j$, let
us try to understand this also from the entropy function
(\ref{eq:EntropyFtn}). Since $(-1)^j=e^{\pi ij}$,
one finds that
\begin{equation}
  (-1)^j\Omega_j\sim
  \exp\left[\frac{N^2(\frac{2\pi i}{3}+\frac{2\omega_\ast}{3})^3}{2\omega_\ast^2}
  +\frac{\omega_\ast+\pi i}{3}j+\cdots\right]+c.c.\ ,
\end{equation}
where $\cdots$ are possible subleading corrections in small
$\frac{1}{N^2}$ and $\frac{1}{j}$ that have not been computed to date.
From this, one obtains
\begin{equation}
  (-1)^j\Omega_j\sim
  \exp\left[{\rm Re}(S(j))+\cdots\right]
  \cos\left[{\rm Im}(S(j))+\pi j+\cdots\right]\ .
\end{equation}
Although the subleading corrections to ${\rm Re}(S(j))$ will not affect
our studies below, the corrections to ${\rm Im}(S(j))$ will be somewhat
important since they will make a finite phase shift of the oscillation.
The corresponding entropy (\ref{entropy-subleading})
improving the black curve of Fig.~\ref{fig:U5U6}
is shown in Fig.~\ref{fig:U5U6bump}.

Firstly, Fig.~\ref{fig:U5U6bump} clearly shows that the
signs of $(-1)^j\Omega_j$ are equal to the sign of
$\cos\left[{\rm Im}(S(j))+\pi j+\eta\right]$. 
As mentioned in the previous paragraph, we empirically fitted the possible 
subleading correction $\eta$ by an $\mathcal{O}(1)$ constant. 
Although $\eta$ is in principle a function of $j$, $N^2$ such as 
$\eta({\textstyle \frac{j}{N^2}})$, constant $\eta$ seems to be reasonably good
within the relatively short ranges of charges in Fig.~\ref{fig:U5U6bump}.\footnote{We also note that,
upon including the 1-loop determinant factor of the Legendre transformation 
(\ref{eq:EntropyFtn}) in this framework, one obtains much better agreements
than those in Fig.~\ref{fig:U5U6bump}. However, we do not show these results 
here since they do not seem to be based on a systematic calculus of the subleading 
terms.} The agreements in Fig.~\ref{fig:U5U6bump} justify our interpretation 
that the oscillation caused by the
complex saddle point accounts for the sign oscillations of $\Omega_j$.
Moreover, Fig.~\ref{fig:U5U6bump} shows that the oscillation
of $\left|\cos\left[{\rm Im}(S(j))+\pi j+\eta\right]\right|$
accounts for the subleading oscillations of our numerically computed
$\log |\Omega_j|$. Therefore, we find that our finite $N$
numerical data strongly supports the detailed structures of the
macroscopic entropy computed at the complex saddle points of
Legendre transformation.

As mentioned in the introduction, it seems that
our interpretation for the complex saddle point 
is very universal. To confirm this expectation,
it will be helpful to study other index-like generating
functions which are simpler than the large $N$ index of
the $\mathcal{N}=4$ Yang-Mills theory. In particular,
for the Yang-Mills index, note that the analytic form of $S(j)$ is known
only to the leading order in large $N$ and $j$.
Due to this limitation, we added an empirical constant $\eta$
at a subleading order to see if the structures of
$S(j)$ and $\Omega_j$ are compatible with each other.
So it will be desirable to study simpler examples in which we can easily
compute the subleading corrections for the precision tests.

\begin{figure}[t!]
    \centering
    \includegraphics[width=0.95\textwidth]{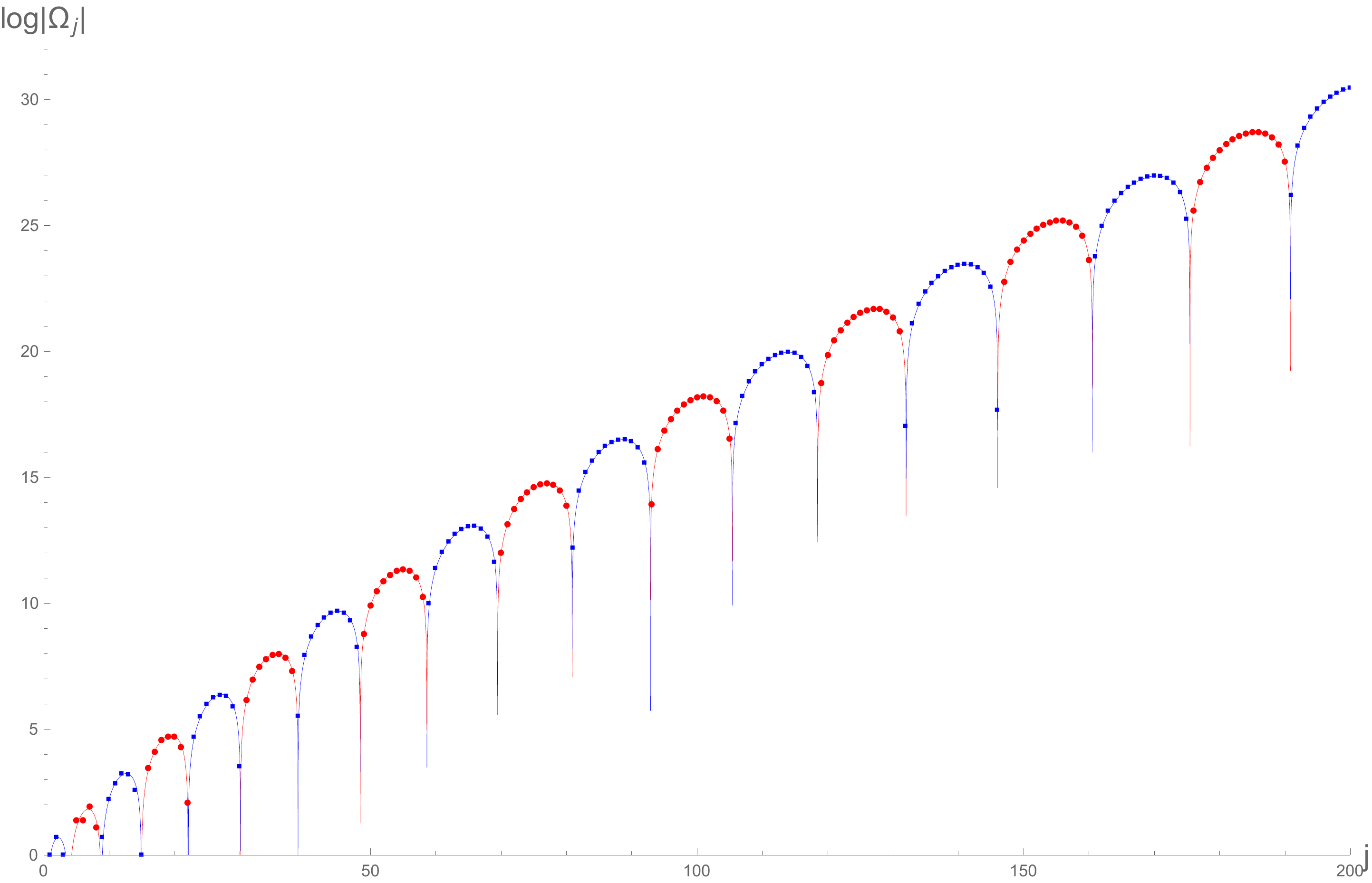}
    \caption{Two plots of $\log |\Omega_j|$ for the MacMahon function.
    Red/blue colors denote the positive/negative signs of $\Omega_j$.}\label{mac}
\end{figure}

As a simple example, consider the inverse of the MacMahon function,
\begin{equation}
  f(x)=\prod_{n=1}^\infty(1-x^n)^n=
  \exp\left[-\sum_{n=1}^\infty\frac{1}{n}
  \frac{x^n}{(1-x^n)^2}\right]\equiv
  \sum_{j=0}^\infty \Omega_j x^j\ .
\end{equation}
Numerically, one can easily expand $f(x)$ in power series of $x$ with
a computer to very high orders. At large charge $j$, one can see that
the resulting $\Omega_j$'s become macroscopic with sign oscillations.
We shall now make an analytic evaluation of the asymptotic entropy
at $j\gg 1$, with necessary subleading corrections in $\frac{1}{j}$
included. We would like to compute
\begin{equation}\label{laplace-macmahon}
  \Omega_j=\frac{1}{2\pi i}\oint \frac{dx}{x}
  x^{-j}f(x)=\frac{1}{2\pi i}\oint \frac{dx}{x}
  \exp\left[j\beta-\sum_{n=1}^\infty\frac{1}{n}
  \frac{e^{-n\beta}}{(1-e^{-n\beta})^2}\right]
\end{equation}
where $x\equiv e^{-\beta}$. The saddle point values $\beta_\ast$ 
of $\beta$ will be small complex numbers with 
${\rm Re}(\beta_\ast)>0$. At small $\beta$, one can use
\begin{equation}
  -\sum_{n=1}^\infty\frac{1}{n}\frac{e^{-n\beta}}{(1-e^{-n\beta})^2}
  =-\frac{\zeta(3)}{\beta^2}-\frac{1}{12}\log \beta-\zeta^\prime(-1,0)
  +\frac{\beta^2}{2880}+\frac{\beta^4}{725760}
  +\frac{\beta^6}{43545600}+\cdots\ ,
\end{equation}
where $\zeta(s)$ is the Riemann zeta function, and
$\zeta^\prime(-1,0)\approx -0.165421$ is the derivative
$\zeta^\prime(s,q)\equiv \frac{\partial\zeta(s,q)}{\partial s}$
of the Hurwitz zeta function. Using this formula with higher order
corrections in small $\beta$, one can approximate the integral
(\ref{laplace-macmahon}) with subleading corrections in
$\frac{1}{j}$ included.
One finds that the following mutually complex conjugate pair of
saddle points are dominant:
\begin{equation}
  \beta_\ast^\pm=e^{\pm\frac{\pi i}{3}}
  \left(\frac{2\zeta(3)}{j}\right)^{\frac{1}{3}}+\frac{1}{36j}
  +\frac{e^{\mp\frac{\pi i}{3}}}{1296(2\zeta(3)j^5)^{\frac{1}{3}}}
  +\cdots\ .
\end{equation}
Performing the Gaussian approximations at these two saddle points
(with some subleading terms included) and adding the two contributions,
one obtains
\begin{eqnarray}
  \Omega_j&\sim&
  \frac{1}{(2\pi)^{\frac{1}{2}}}
  \sum_{\pm}\exp\left[\frac{3}{2}e^{\pm\frac{\pi i}{3}}
  (2\zeta(3) j^2)^{\frac{1}{3}}
  +\frac{1}{36}\log j-\zeta^\prime(-1,0)
  -\frac{\log(2\zeta(3))}{36}\mp\frac{\pi i}{36}+\cdots\right]\\
  &&\hspace{1cm}
  \times\left[3e^{\mp\frac{\pi i}{3}}
  \left(\frac{j^4}{2\zeta(3)}\right)^{\frac{1}{3}}
  +\frac{1}{4}e^{\pm\frac{\pi i}{3}}
  \left(\frac{j}{2\zeta(3)}\right)^{\frac{2}{3}}
  -\frac{1}{216\zeta(3)}+\cdots\right]^{-\frac{1}{2}}
  \cdot\left[1+\cdots\frac{}{}\!\!\right]
  \nonumber\ .
\end{eqnarray}
Here, the three factors on the right hand side come from 
the saddle point action, the 1-loop determinant, and 
possible higher loop corrections, respectively.
We plot this asymptotic $\log |\Omega_j|$ in Fig.~\ref{mac}, together
with the dotted plot obtained from the series expansion up to
$\mathcal{O}(x^{200})$ order.

\vspace{\baselineskip}

\noindent{\bf\large Acknowledgements}

\noindent
We thank Ashoke Sen for suggesting us to study the finite $N$ indices to
better understand AdS black holes, which was the starting point of this project.
We also thank Shota Komatsu and Xi Yin for the helpful comments and questions.
PA, SC and SK are supported in part by the National Research Foundation of Korea (NRF) Grant 2018R1A2B6004914. The work of PA is also supported in part by the Korea Research Fellowship Program through the National Research Foundation of Korea funded by the Ministry of Science, ICT and Future Planning, grant number 2016H1D3A1938054 and in part by the Royal Society Research Fellows Enhancement Award, grant no. RGF\textbackslash EA\textbackslash 181049. The work of SC is also
supported by NRF-2017-Global Ph.D. Fellowship Program. JK is supported by the NSF
grant PHY-1911298. JN is supported by a KIAS Individual Grant PG76401.

\noindent

\appendix
\section{Numerical data}
\begin{figure}[h!]
    \centering
    \includegraphics[height=8cm]{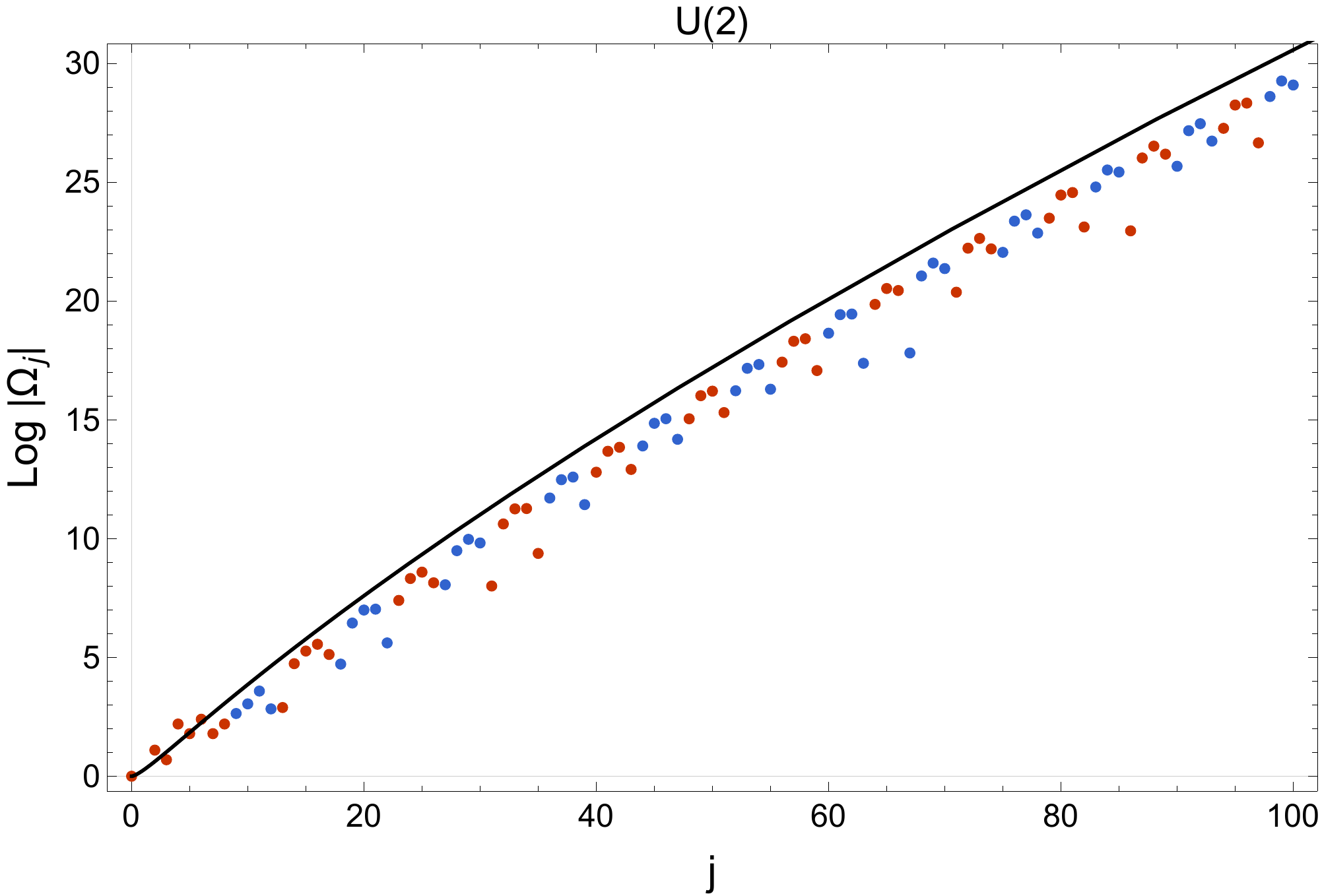}
    \caption{$N=2$}
    \label{fig:U2}
\end{figure}

\begin{figure}[h!]
    \centering
    \includegraphics[height=8cm]{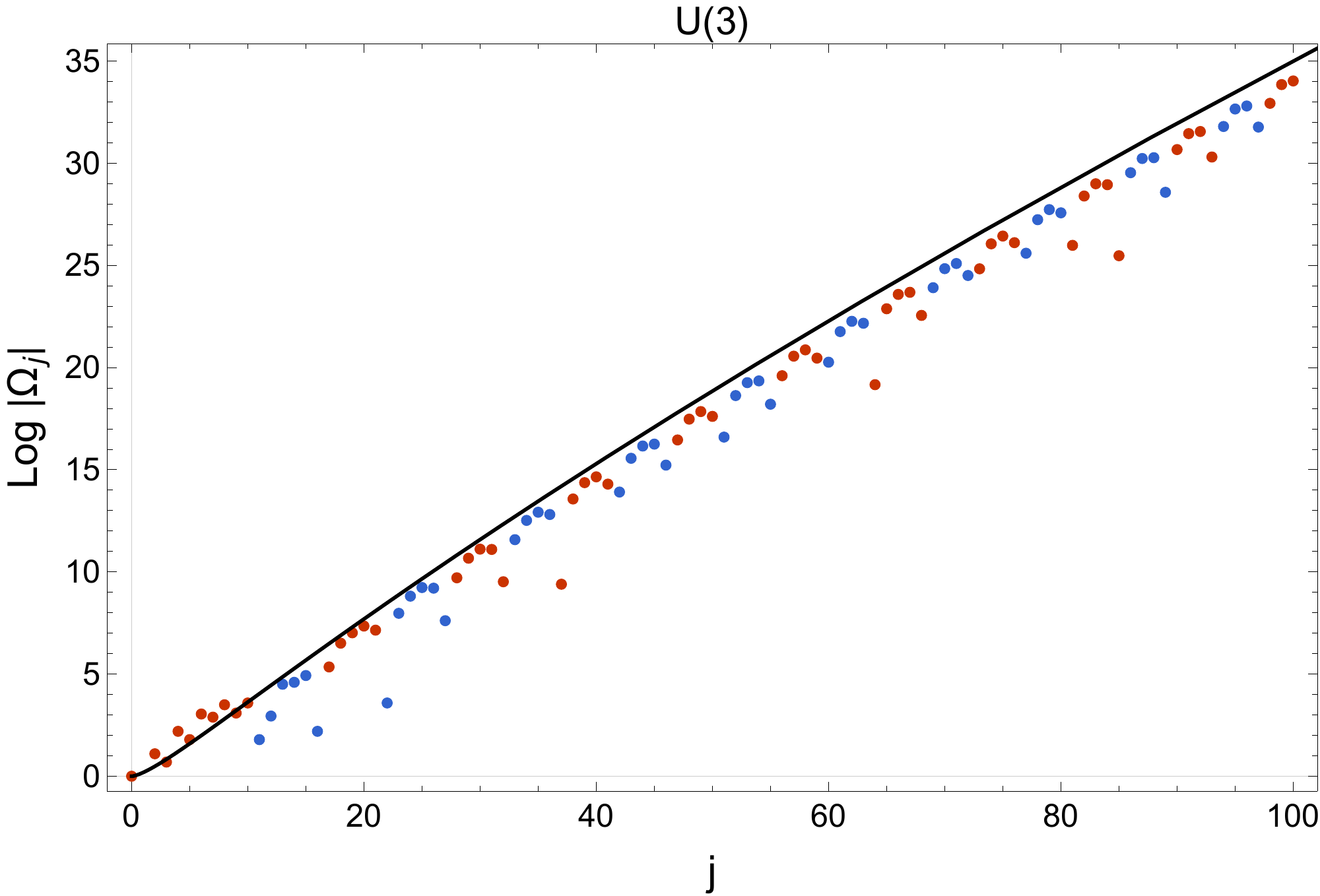}
    \caption{$N=3$}
    \label{fig:U3}
\end{figure}

\begin{figure}[h!]
    \centering
    \includegraphics[height=8cm]{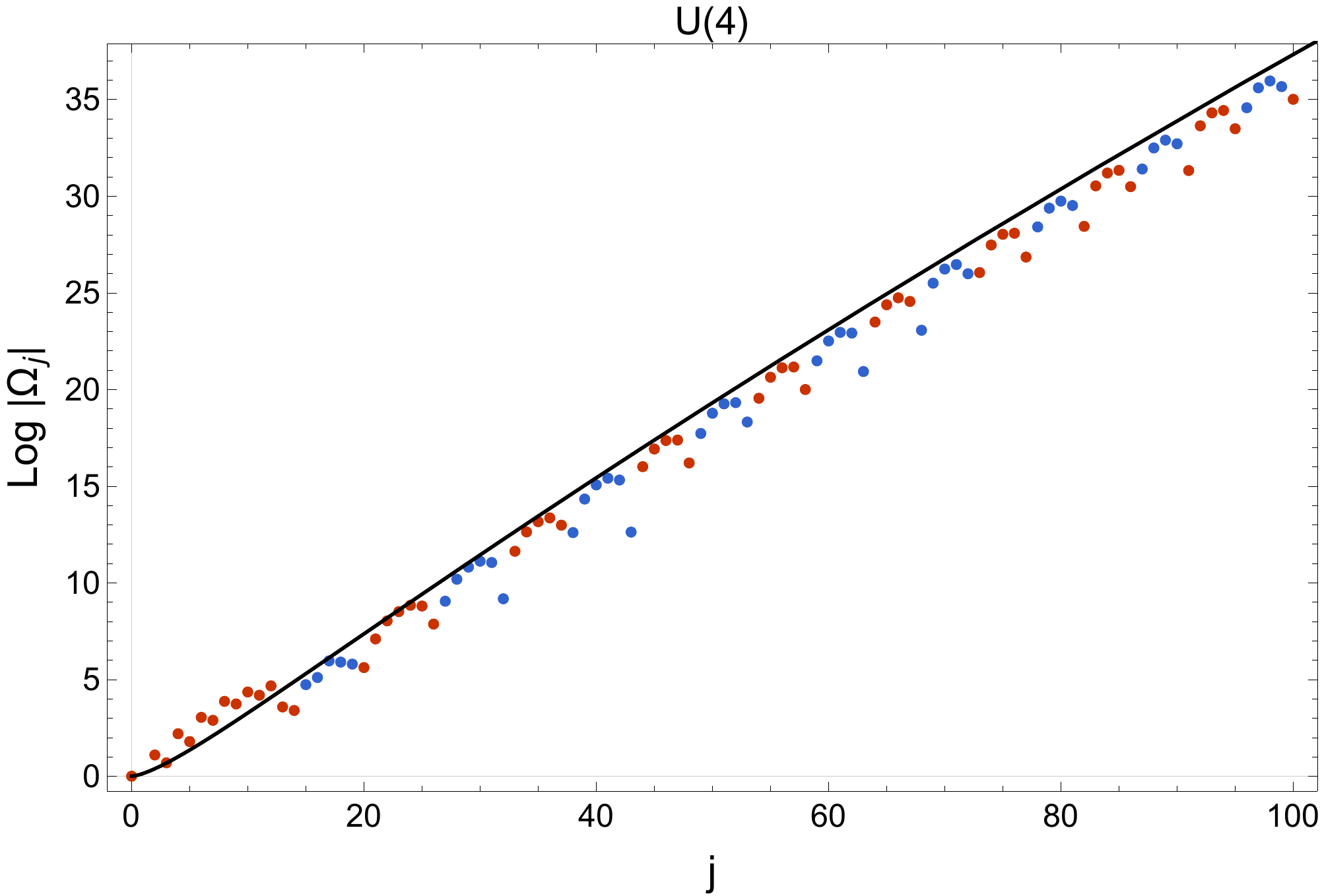}
    \caption{$N=4$}
    \label{fig:U4}
\end{figure}

Here we collect the numerical expression for the $U(N)$ index with $2 \leq N \leq 5$. The following expression is the $U(2)$ index, which is illustrated in Figure~\ref{fig:U2}.
\begin{align}
    \label{eq:u2index}
    Z_{U(2)} &= 1+3 x^2-2 x^3+9 x^4-6 x^5+11 x^6-6 x^7+9 x^8+14 x^9-21 x^{10}+36 x^{11}-17 x^{12} \nonumber \\&
    -18 x^{13}+114 x^{14}-194 x^{15}+258 x^{16}-168 x^{17}-112
   x^{18}+630 x^{19}-1089 x^{20}\nonumber \\&
   +1130 x^{21}-273 x^{22}-1632 x^{23}+4104 x^{24}-5364 x^{25}+3426 x^{26}+3152 x^{27}\nonumber \\&
   -13233 x^{28}+21336 x^{29}-18319 x^{30}-2994 x^{31}+40752 x^{32}-76884 x^{33}+78012 x^{34}\nonumber \\&
   -11808 x^{35}-121384 x^{36}+262206 x^{37}-293145 x^{38}+91904
   x^{39}+359775 x^{40}\nonumber \\&
   -867906 x^{41}+1026540 x^{42}-404454 x^{43}-1086312 x^{44}+2815744 x^{45}-3415932 x^{46}\nonumber \\&
   +1436112 x^{47}+3403791
   x^{48}-9007578 x^{49}+10895604 x^{50}-4420644 x^{51}-11068260 x^{52}\nonumber \\&
   +28481682 x^{53}-33440475 x^{54}+11822670 x^{55}+36950502
   x^{56}-88878842 x^{57}\nonumber \\&
   +98770059 x^{58}-25918986 x^{59}-124747447 x^{60}+272655942 x^{61}-279580701 x^{62}\nonumber \\&
   +35207790 x^{63}+419441625
   x^{64}-818211192 x^{65}+751976333 x^{66}+54317328 x^{67}\nonumber \\&
   -1386833514 x^{68}+2387940758 x^{69}-1893048381 x^{70}-700663056 x^{71}+4467470232
   x^{72}\nonumber \\&
   -6731222448 x^{73}+4333120557 x^{74}+3746183998 x^{75}-13926217512 x^{76}+18169226454 x^{77}\nonumber \\&
   -8426843619 x^{78}-15799669950
   x^{79}+41774162736 x^{80}-46405515308 x^{81}\nonumber \\&
   +10894454985 x^{82} +58624684746 x^{83}-119915881179 x^{84}+110030518596 x^{85}\nonumber \\&
   +9268878210 x^{86}-198813575484 x^{87}+327212977320 x^{88}-233510264916 x^{89}\nonumber \\&
   -140308837617 x^{90}+626333831526 x^{91}-840626319591 x^{92}+404682823524 x^{93}\nonumber \\&
   +692617559553 x^{94} -1844851526580 x^{95}+2001340988797 x^{96}-375443664666 x^{97}\nonumber \\&
   -2639017467255 x^{98}
   +5082041971496 x^{99}-4283590699023 x^{100}+ \mathcal{O}(x^{101})
\end{align}
The next expression is the $U(3)$ index, whose $\log|\Omega_j|$ is drawn in Figure~\ref{fig:U3}.
\begin{align}
    \label{eq:u3index}
    Z_{U(3)} &= 1+3 x^2-2 x^3+9 x^4-6 x^5+21 x^6-18 x^7+33 x^8-22 x^9+36 x^{10}+6 x^{11}-19 x^{12}+90 x^{13}\nonumber \\&
    -99 x^{14}+138 x^{15}-9 x^{16}-210 x^{17}+672
   x^{18}-1116 x^{19}+1554 x^{20}-1270 x^{21}-36 x^{22}\nonumber \\&
   +2898 x^{23}-6705 x^{24}+10224 x^{25}-9918 x^{26}+2018 x^{27}+16470 x^{28}-42918 x^{29}\nonumber \\&
   +66906 x^{30}-66006 x^{31}+13566 x^{32}+106404 x^{33}-273204 x^{34}+407442 x^{35}-364710 x^{36}\nonumber \\&
   -12024 x^{37}+778272 x^{38}-1731542 x^{39}+2300499 x^{40}-1611774 x^{41}-1093848 x^{42}\nonumber \\&
   +5702562 x^{43}-10400586 x^{44}+11407626 x^{45}-4086693 x^{46}-13996782 x^{47}\nonumber \\&
   +38712766 x^{48}-56127654 x^{49}+44316099 x^{50}+16085226 x^{51}-122617179 x^{52}\nonumber \\&
   +231054624 x^{53}-251544720 x^{54}+80412606 x^{55}+324099348
   x^{56}-844286204 x^{57}\nonumber \\&
   +1147990887 x^{58}-767030682 x^{59}-628392075 x^{60}+2808255348 x^{61}-4642468821 x^{62}\nonumber \\&
   +4223264234 x^{63}+209141406 x^{64}-8584019040 x^{65}+17327115906 x^{66}\nonumber \\&
   -19194283332 x^{67}+6197598675 x^{68}+24052600650 x^{69}-61026825105 x^{70}\nonumber \\&
   +78594793644 x^{71}-43722790228 x^{72}-60628872366 x^{73}+205754044713 x^{74}\nonumber \\&
   -300949636742 x^{75}+217767461283 x^{76}+129914189388 x^{77}-671070962823
   x^{78}\nonumber \\&
   +1099745830260 x^{79}-937888762842 x^{80}-191081792160 x^{81}+2135620393074 x^{82}\nonumber \\&
   -3884644088484 x^{83}+3715774679244
   x^{84}-114903322902 x^{85}-6683223253806 x^{86}\nonumber \\&
   +13381744369680 x^{87}-13925733216507 x^{88}+2562254228766 x^{89}\nonumber \\&
   +20719792872015 x^{90}-45245335312008 x^{91}+50127612882930 x^{92}\nonumber \\&
   -14402257204784 x^{93}-64103402035710 x^{94} +150872971344750 x^{95} \nonumber \\&
   -174917721819708 x^{96}+62316941736600 x^{97} +199276922573595 x^{98}\nonumber \\&
   -497907763520398 x^{99}+595172510765379 x^{100}+ \mathcal{O}(x^{101})
\end{align}
The $U(4)$ index comes next. The corresponding figure, $\log|\Omega_j|$ vs $j$, is drawn in Figure~\ref{fig:U4}.
\begin{align}
    \label{eq:u4index}
    Z_{U(4)} &= 1+3 x^2-2 x^3+9 x^4-6 x^5+21 x^6-18 x^7+48 x^8-42 x^9+78 x^{10}-66 x^{11}+107 x^{12}\nonumber \\&
    -36 x^{13}+30 x^{14}+114 x^{15}-165 x^{16}+390 x^{17}-366
   x^{18}+330 x^{19}+276 x^{20}-1212 x^{21}\nonumber \\&
   +3081 x^{22}-4986 x^{23}+6924 x^{24}-6654 x^{25}+2616 x^{26}+8528 x^{27}-26571 x^{28}\nonumber \\&
   +49800 x^{29}-67651 x^{30}+63096 x^{31}-9678 x^{32}-112980 x^{33}+307098 x^{34}-522066 x^{35}\nonumber \\&
   +634029 x^{36}-436260 x^{37}-296460 x^{38}+1682020
   x^{39}-3497613 x^{40}+4937946 x^{41}\nonumber \\&
   -4501122 x^{42}+304512 x^{43}+8971113 x^{44}-22380734 x^{45}+34738953 x^{46}\nonumber \\&
   -35553996 x^{47}+10888602 x^{48}+49956294 x^{49}-142303191 x^{50}+231744000 x^{51}\nonumber \\&
   -246464136 x^{52}+90402078 x^{53}+309123032 x^{54}-917051802 x^{55}+1494916050 x^{56}\nonumber \\&
   -1558557796 x^{57}+485393061 x^{58}+2144544540 x^{59}-5983505013 x^{60}\nonumber \\&
   +9333423798 x^{61}-9004631841 x^{62}+1231871108
   x^{63}+15915475365 x^{64}\nonumber \\&
   -38937814944 x^{65}+55770600072 x^{66}-46223256036 x^{67}-10405285128 x^{68}\nonumber \\&
   +118932061824 x^{69}-247095009891
   x^{70}+311970699564 x^{71}-193686936205 x^{72}\nonumber \\&
   -205315072914 x^{73}+855723695370 x^{74}-1490314195506 x^{75}+1572823900839 x^{76}\nonumber \\&
   -458786822988 x^{77}-2181976709955 x^{78}+5759182587780 x^{79}-8289856609587 x^{80}\nonumber \\&
   +6601945579040 x^{81}+2245784042823
   x^{82}-18254661918174 x^{83}+35440988310091 x^{84}\nonumber \\&
   -40697268408630 x^{85}+17515834035681 x^{86}+43558153249536 x^{87}\nonumber \\&
   -129719118983523 x^{88}+194052483593046 x^{89}-160650745697554 x^{90}\nonumber \\&
   -40311995227758 x^{91}+407070606690366 x^{92}-795660945732754 x^{93}\nonumber \\&
   +899816226757623 x^{94}-349806028105302 x^{95}-1035026648995290 x^{96}\nonumber \\&
   +2903482927460364 x^{97}-4145273582018487 x^{98}+3091519137195862
   x^{99}\nonumber \\&
   +1604158693277994 x^{100} + \mathcal{O}(x^{101})
\end{align}
The following series expression is the $U(5)$ index. The relevant plot of $\log|\Omega_j|$ is given in Figure~\ref{fig:U5}.
\begin{align}
    \label{eq:u5index}
    Z_{U(5)} &= 1+3 x^2-2 x^3+9 x^4-6 x^5+21 x^6-18 x^7+48 x^8-42 x^9+99 x^{10}-96 x^{11}+172 x^{12}\nonumber \\&
    -156 x^{13}+252 x^{14}-160 x^{15}+195 x^{16}+48 x^{17}-127
   x^{18}+612 x^{19}-783 x^{20}+1258 x^{21}\nonumber \\&
   -948 x^{22}+450 x^{23}+1921 x^{24}-5430 x^{25}+11793 x^{26}-18812 x^{27}+26379 x^{28}-27750 x^{29}\nonumber \\&
   +17809 x^{30}+15648 x^{31}-78324 x^{32}+175030 x^{33}-285576 x^{34}+366024 x^{35}-323807 x^{36}\nonumber \\&
   +38856 x^{37}+624894 x^{38}-1718016
   x^{39}+3094992 x^{40}-4226862 x^{41}+4098270 x^{42}\nonumber \\&
   -1210728 x^{43}-5968935 x^{44}+18061488 x^{45}-33152565 x^{46}+44941584 x^{47}-41448422
   x^{48}\nonumber \\&
   +6241896 x^{49}+75761478 x^{50}-205993284 x^{51}+354209109 x^{52}-440168670 x^{53}\nonumber \\&
   +328572109 x^{54}+142704804 x^{55}-1079522706
   x^{56}+2385844062 x^{57}-3584202447 x^{58}\nonumber \\&
   +3694263972 x^{59}-1331772481 x^{60}-4771857420 x^{61}+14697077445 x^{62}-25833114276
   x^{63}\nonumber \\&
   +31549909440 x^{64}-21264664440 x^{65}-16439430686 x^{66}+86286819246 x^{67}\nonumber \\&
   -174750537792 x^{68}+238416590234 x^{69}-201108631665
   x^{70}-27442949994 x^{71}\nonumber \\&
   +499854484406 x^{72}-1146580228470 x^{73}+1684959423831 x^{74}-1584800711048 x^{75}\nonumber \\&
   +184556608692
   x^{76}+2953939765242 x^{77}-7447464688605 x^{78}+11432006505378 x^{79}\nonumber \\&
   -11287805022885 x^{80}+2416173603110 x^{81}+18314405974467
   x^{82}-48439160197746 x^{83}\nonumber \\&
   +75397207473690 x^{84}-74801457474012 x^{85}+16057846263102 x^{86}+120661512888900 x^{87}\nonumber \\&
   -316568078311605 x^{88}+485306430414990 x^{89}-464824039417731 x^{90}\nonumber \\&
   +60350744120262 x^{91}+837845036799732 x^{92}-2071759782098082 x^{93}\nonumber \\&
   +3041713804417725 x^{94}-2691482911939584 x^{95}-156200831519985 x^{96}\nonumber \\&
   +5991608828442690 x^{97}-13462930267605216 x^{98}+18424199416716136
   x^{99}\nonumber \\&
   -14187219139048212 x^{100}+ \mathcal{O}(x^{101})
\end{align}


\pagebreak
\providecommand{\href}[2]{#2}\begingroup\raggedright\endgroup

\end{document}